\documentclass[aps,prb,preprint,superscriptaddress,showpacs,amsmath]{revtex4}

\usepackage[dvips]{graphicx,epsfig}
\usepackage{natbib}

\begin{document}

\title{Charge control in laterally coupled double quantum dots}

\author{G. Mu\~{n}oz-Matutano}
\affiliation{ICMUV, Instituto de Ciencia de Materiales, Universidad de Valencia, P.O. Box 22085, 46071 Valencia, Spain.}

\author{M. Royo}
\affiliation{Departament de Qu\'{i}mica F\'{i}sica i Anal\'{i}tica, Universitat Jaume I, Castell\'{o}n, Spain.}

\author{J. I. Climente}
\affiliation{Departament de Qu\'{i}mica F\'{i}sica i Anal\'{i}tica, Universitat Jaume I, Castell\'{o}n, Spain.}

\author{J. Canet-Ferrer}
\affiliation{ICMUV, Instituto de Ciencia de Materiales, Universidad de Valencia, P.O. Box 22085, 46071 Valencia, Spain.}

\author{D. Fuster}
\affiliation{ICMUV, Instituto de Ciencia de Materiales, Universidad de Valencia, P.O. Box 22085, 46071 Valencia, Spain.}

\author{P. Alonso-Gonz\'{a}lez}
\affiliation{IMM, Instituto de Microelectr\'{o}nica de Madrid (CNM, CSIC), Isaac Newton 8,
 28760 Tres Cantos, Madrid, Spain.}

\author{I. Fern\'{a}ndez-Mart\'{\i}nez}
\affiliation{IMM, Instituto de Microelectr\'{o}nica de Madrid (CNM, CSIC), Isaac Newton 8,
 28760 Tres Cantos, Madrid, Spain.}

\author{J. Mart\'{\i}nez-Pastor}
\affiliation{ICMUV, Instituto de Ciencia de Materiales, Universidad de Valencia, P.O. Box 22085, 46071 Valencia, Spain.}

\author{Y. Gonz\'{a}lez}
\affiliation{IMM, Instituto de Microelectr\'{o}nica de Madrid (CNM, CSIC), Isaac Newton 8,
 28760 Tres Cantos, Madrid, Spain.}

\author{L. Gonz\'{a}lez}
\affiliation{IMM, Instituto de Microelectr\'{o}nica de Madrid (CNM, CSIC), Isaac Newton 8,
 28760 Tres Cantos, Madrid, Spain.}

\author{F. Briones}
\affiliation{IMM, Instituto de Microelectr\'{o}nica de Madrid (CNM, CSIC), Isaac Newton 8,
 28760 Tres Cantos, Madrid, Spain.}

\author{B. Al\'{e}n}
\email[]{benito.alen@imm.cnm.csic.es}
\affiliation{IMM, Instituto de Microelectr\'{o}nica de Madrid (CNM, CSIC), Isaac Newton 8,
 28760 Tres Cantos, Madrid, Spain.}

\date{\today}

\begin{abstract}
We investigate the electronic and optical properties of InAs double quantum dots grown on GaAs (001) and laterally aligned along the [110]
crystal direction. The emission spectrum has been investigated as a function of a lateral electric field applied along the quantum dot pair
mutual axis. The number of confined electrons can be controlled with the external bias leading to sharp energy shifts which we use to
identify the emission from neutral and charged exciton complexes. Quantum tunnelling of these electrons is proposed to explain the reversed
ordering of the trion emission lines as compared to that of excitons in our system.
\end{abstract}


\maketitle

Optical spin initialization of individual electrons is a fundamental resource for quantum information science which relies on our ability
to control the charge in a quantum dot (QD) molecule.~\cite{Robledo2008,Kim2008,Kim2011} In the last years, vertically aligned QDs have
been fabricated with great success. In these systems, exciton coupling signatures including energy anticrossings of neutral and charged
exciton complexes have been demonstrated by applying an electric field in the growth
direction.~\cite{Krenner2005,Stinaff2006,Scheibner2007} Lateral QD molecules would be a better candidate for scaling-up the electronic
coupling from two to several QDs applying individual lateral gates. Previous demonstrations of electronic coupling in a lateral QD pair
have been based on analysis of the anomalous Stark shifts and photon correlation statistics of the neutral exciton under a lateral electric
field.~\cite{Beirne2006} Yet, the observation of electrically tunable energy anticrossings in lateral QD molecules remains a difficult task
due the exponential decrease of the tunnel coupling energy with the center to center QD distance, $d$.~\cite{Beirne2006,Wang2009,Peng2010}
In the following, we investigate the emission spectrum of electrically tunable lateral QD pairs with varying number of electrons. For
typical inter-dot distances $d\sim30-40$ nm, we find that electron tunnelling phenomena affect the negative trion emission energy before
clear exciton anticrossings may take place.

\begin{figure}[htb]
\begin{center}
\includegraphics[width=80mm]{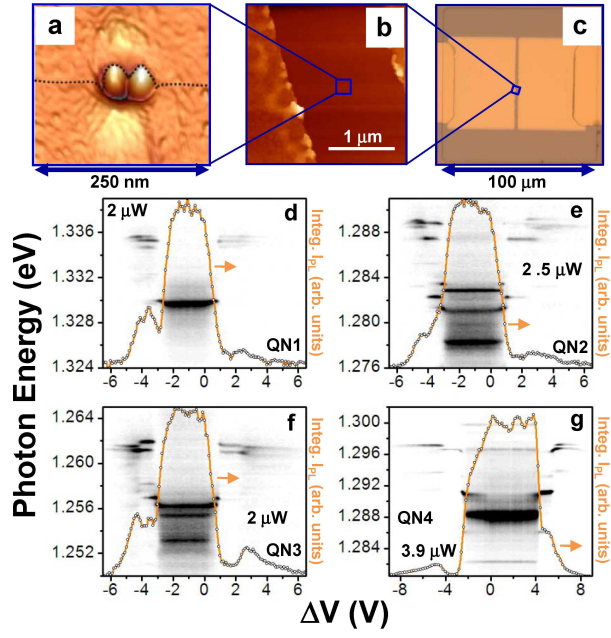}%
\caption{B. Al\'en et al.} \label{Fig1}
\end{center}
\end{figure}

For the present study, QD pairs aligned in the [110] crystal direction were fabricated on GaAs nanoholes using a modified droplet epitaxy
growth procedure as described in detail elsewhere.~\cite{Alonso-Gonzalez2009b} The nanostructures were grown on a 0.5-$\mu$m-thick undoped
GaAs buffer layer and capped by 100 nm of undoped GaAs. Atomic force micrographs (AFM) performed on a similar uncapped sample revealed that
each QD in the pair has slightly different height, with average values of $5.3 \pm 0.9$ nm and $6.6 \pm 1.5$ nm, respectively, and center
to center separation equal to their average diameter $37 \pm 4$ nm [Fig.~\ref{Fig1}(a)]. The morphological analysis also revealed that QD
pair formation occurs in 95 \% of the cases with a small probability for single QDs or empty nanoholes. The low areal density of $2\times
10^8$ cm$^{-2}$ is adequate to study individual quantum nanostructures.

To apply an electric field along the QD pair mutual axis, we defined metal-semiconductor-metal (MSM) diodes by evaporation of two metal
contacts (15 nm Mo + 30 nm Au) on top of a 100 $\mu$m square mesas. The contacts are separated by a 80 $\mu$m long$\times$1.5 $\mu$m wide
undoped GaAs channel embedding the nanostructures as shown in Figures~\ref{Fig1}(a-c). The micro-photoluminescence ($\mu$-PL) of individual
QD pairs was collected at 5 K using a fiber based confocal microscope, excited with 785 nm continuous wave laser light, dispersed by a
2$\times$0.3 m focal length double spectrograph and detected with a peltier cooled Silicon CCD camera. The spectral resolution of our setup
is $\sim$90 $\mu$eV full width at half maximum (FWHM).

In the last few years, several groups have investigated the emission of single semiconductor nanostructures in the presence of a lateral
electric field. For moderate electric fields, or when the separation between the contacts is large, the changes observed in the QD spectrum
have been related to the modulation of the carrier capture probability induced by the external field.~\cite{Moskalenko2006,Moskalenko2007}
The capture mechanisms also play the most important role in single QDs dynamically driven by surface acoustic
waves.~\cite{Volk2010,Volk2011} The laterally applied bias can also modulate the electronic confinement levels. This requires of larger
electric fields or smaller contact separation for a given bias range. In this regime, the exciton wavefunction can be directly modified
leading to energy shifts, carrier tunnelling and fine structure splitting reduction among other
effects.~\cite{Kowalik2005,Stavarache2006,Gerardot2007,Reimer2008,Alen2009} With a channel width of only 1.5 $\mu$m, our MSM diodes have
been designed to apply large electric fields in the [110] crystal direction (0-60 kV/cm). This is required to tune independently the
exciton energy of the two QDs in a lateral molecule and, if their separation were small enough, to observe resonant quantum tunnelling
phenomena.

The contour maps in fig.~\ref{Fig1}~(d-f) show the evolution of the $\mu$-PL spectrum as a function of the lateral bias, $\Delta V$, for
four different nanostructures (QN1-QN4). In each case, the evolution of the spectrally integrated intensity is also drawn [orange (spotted)
lines]. The integrated intensity remains constant within 10\% for a broad voltage range around a certain bias $\Delta V_c\neq0$ V and
diminishes down to zero for larger positive or negative bias. While the intensity decreases, the emission spectrum also changes giving rise
to blue-shifted spectral features which are different in each nanostructure as shown in the $\mu$-PL contour maps. These emission patterns
can be examined to distinguish between single quantum dots (QN1) and double quantum dots (QN2-QN4) as exemplarily explained below for QN1
and QN2.

\begin{figure}[htb]
\begin{center}
\includegraphics[width=70 mm]{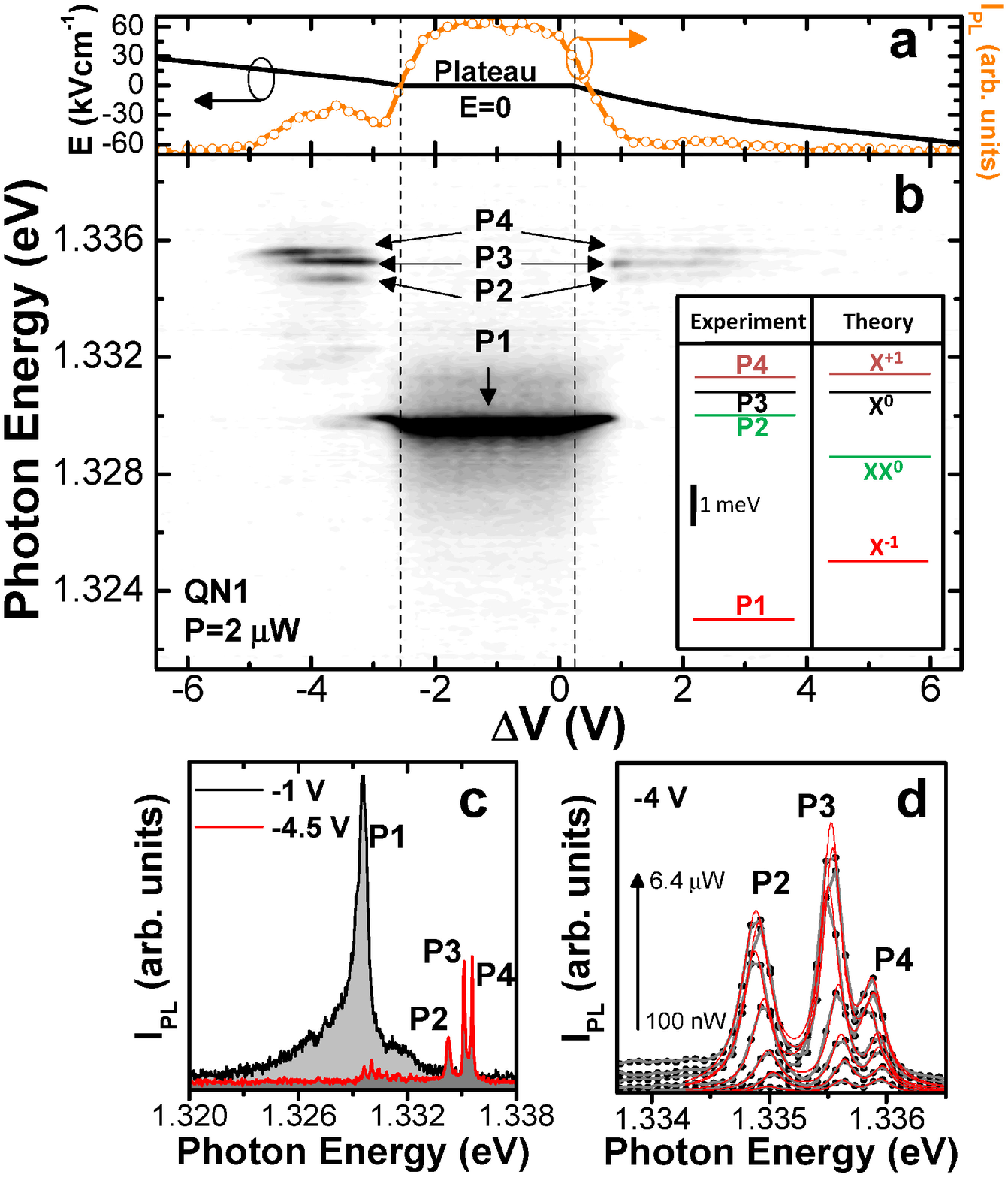}%
\caption{B. Al\'en et al.} \label{Fig2}
\end{center}
\end{figure}

To do so, we have analyzed the device operation using semi-analytical transport equations valid for one-dimensional (1D) MSM
structures.~\cite{endnote2} The simulations are performed in dark conditions and explain why the spectrum is not fully symmetric around
$\Delta V=$0 V. Attending to the particular position of the nanostructure within the GaAs channel, the electric field is calculated
neglecting possible screening effects induced by the photogenerated carriers. Figure~\ref{Fig2}(a) show the $E(\Delta V)$ dependence [black
line] at the position of QN1 which we estimate according to the model at $\sim$530 nm from the left contact. The curve shows a plateau
which extends over the bias region where the field is still zero at this position. In this region, the integrated intensity [orange
(spotted) line] and the $\mu$-PL spectrum [contour plot map in fig.~\ref{Fig2}(b)] is independent of the voltage and, in the case of QN1,
is characterized by a single broad resonance at 1.330 eV (labelled as P1). As we approach to the edges of the plateau raising $|\Delta V|$,
the electric field in the vicinity of the nanostructure also increases. In such situation, the possibility of an enhanced capture of
carriers driven by the external lateral field was discussed by Moskalenko \textit{et al}.~\cite{Moskalenko2007} They found that the overall
QD integrated intensity increased rather than decreased for both positive and negative bias and also reported switching between spectral
lines. The later was explained by the uneven capture of electron and holes and was found strongly dependent on the excitation energy and
power and also on the temperature. In our case, we observe an overall reduction of the integrated intensity and switching between spectral
lines which are similar in a wide range of excitation powers and temperatures.~\cite{endnote2} Our observations should be thus related to
the large electric fields present in our devices and to intrinsic properties of InAs quantum dots and quantum dot pairs grown by modified
droplet epitaxy.

The decrease of the integrated intensity at both sides of the central plateau might be explained through the combination of two effects.
First, electron and holes photogenerated above barrier can drift away from the illuminated area, before being captured in the quantum dots,
contributing to the subnanosecond photocurrent response of MSM photodetectors of this size.~\cite{Wei1981} Secondly, carriers already
confined in the nanostructure can tunnel out due to the large applied field. The later causes the switching of spectral lines which we
associate to the recombination of exciton complexes with varying number of electrons.~\cite{Alen2005} If the crystallization of the Ga
droplet is not complete, arsenic vacancies arise during growth creating localized states in the gap. These localized states are close in
energy to the electron confined levels and are occupied by one or more electrons depending on their state of valence.~\cite{Laasonen1992}
Thus, in absence of an electric field, the negatively charged environment leads to a luminescence spectrum dominated by negatively charged
exciton complexes.~\cite{Alonso-Gonzalez2007b} When a bias is applied in either direction, these electrons are swept by the electric field
leading to neutral or positively charged exciton recombination.

Together with the external field, the local field associated to ionized defects~\cite{Kamada2008,Moskalenko2007} and the screening field
created by the accumulation of photogenerated carriers in the metal-semiconductor interfaces~\cite{Wei1981} must be considered. The local
field fluctuates due to the dynamics of the charged environment.~\cite{Kamada2008} This broadens the emission lines by spectral diffusion,
as shown in Figure~\ref{Fig2}(c) for the spectral line P1 (FWHM$\sim$800 $\mu$eV). The fluctuations are largely minimized once the extra
charges have been swept and switching to spectral lines P2-P4 (FWHM$\sim$110-260 $\mu$eV) takes place. Meanwhile, the screening is likely
the responsible of small energy shifts which we observe when varying the excitation power at constant bias [Fig.~\ref{Fig2}(c)]. Both
effects are small, typically 100-500 $\mu$eV, and can be disregarded for the analysis of the charge tuning effect where large shifts (1-8
meV), associated to carrier-carrier interactions, are induced by the external bias. Both effects can be also minimized by using a resonant
excitation scheme.~\cite{Moskalenko2007}

The charge tunability is crucial for applications in quantum information technology and also to identify the different spectral lines in
our experiment. The tunnelling rates are determined by the carrier confinement energies and therefore depend on the QD size and the Coulomb
interactions between electrons and holes.~\cite{Warburton2000,Alen2005} This leads to $\mu$-PL contour maps with characteristic stair-case
patterns and energy splittings which are different for single QDs and QD pairs.

The four spectral lines (P1-P4) in the spectrum of Fig.~\ref{Fig2}(b) can be well described assuming that QN1 is a single QD. To do so, the
electronic structure has been calculated using a 2D effective mass model for electrons and heavy holes.~\cite{endnote2} We calculate the
emission energy of the neutral exciton (X$^0$), negative and positive trions (X$^-$, X$^+$), biexciton (XX$^0$) and negative quarton
(X$^{2-}$) [Inset of Fig.~\ref{Fig2}(b)]. By comparing the energetic ordering with that of the experiment, we find that lines P1,P2,P3 and
P4 correlate well with X$^-$, XX$^0$, X$^0$ and X$^+$ optical transitions, respectively. The X$^{2-}$ triplet resonance, which would show
up at lower energies, is not observed in the spectrum indicating that only one electron is being transferred from the environment to this
particular QD.

\begin{figure}[htb]
\begin{center}
\includegraphics[width=80 mm]{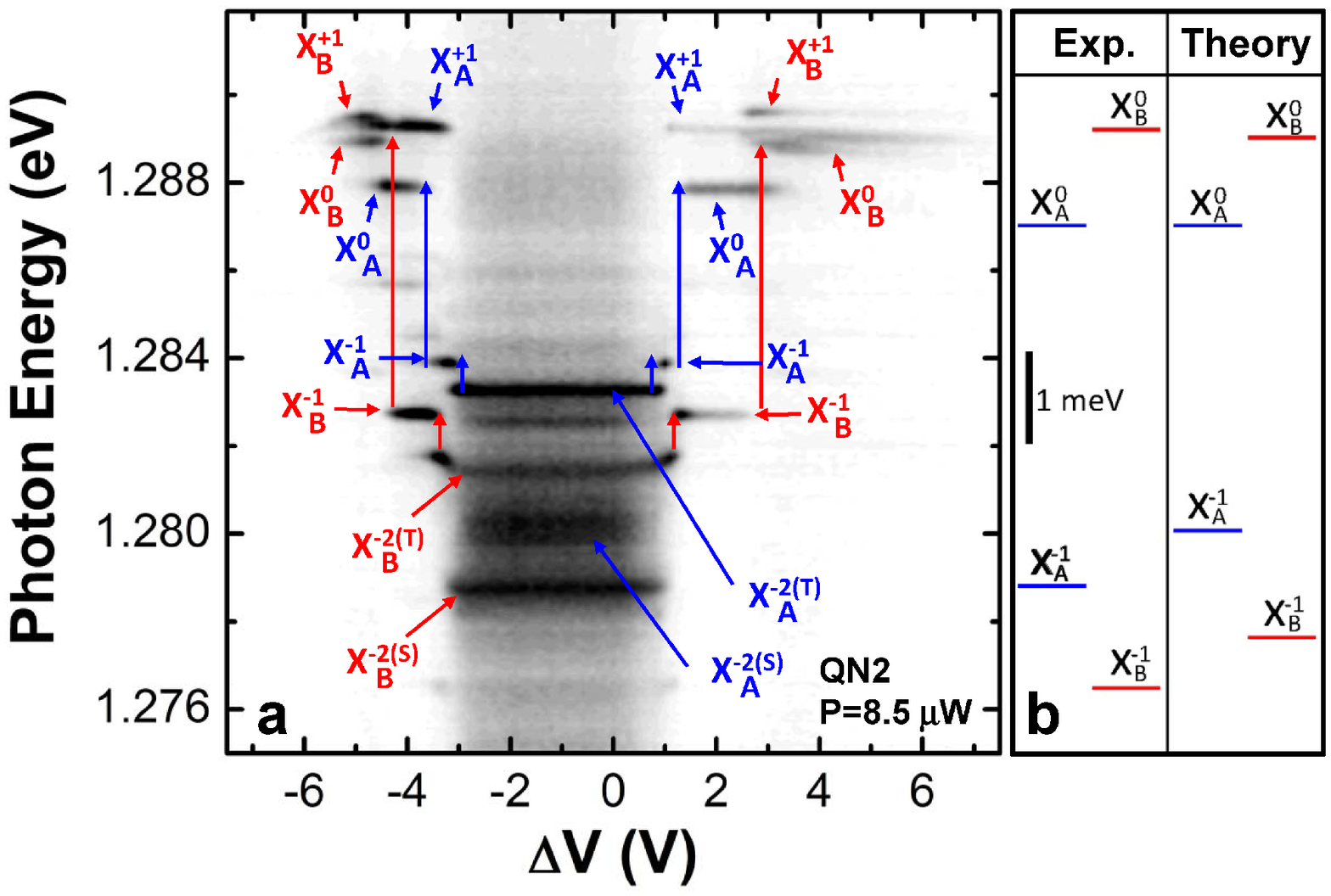}%
\caption{B. Al\'en et al.} \label{Fig3}
\end{center}
\end{figure}

AFM performed in uncapped samples reveals that most of the nanoholes contain lateral QD pairs. Accordingly, most of our spectra can not be
described assuming just a single QD. In Figure~\ref{Fig3}, we analyze in detail the spectrum of QN2 which is characterized by twice the
number of spectral lines expected for a single QD.~\cite{Beirne2006,Pomraenke2008,Abbarchi2009a,Wang2009} The spectrum of QN2 can be
understood by the tentative assignment proposed in the figure, which is roughly that of two single QDs, A and B, with emission from
X$^{2-}$, $X^-$, $X^0$ and $X^+$. The two QDs are asymmetric, as usually observed in AFM, with A being slightly bigger. Starting from the
X$^{2-}$ transitions (singlet and triplet), with increasing external field, the number of additional electrons is tuned from two to zero
and each QD in the pair follows its own Coulomb staircase: X$^{2-}$ $\rightarrow$ X$^-$ $\rightarrow$ X$^0$ $\rightarrow$ X$^+$. This shows
that net confined charge can be controlled in a QD pair by applying a lateral electric field.

The energy ordering in the spectral assignment of Fig.~\ref{Fig3} is the usual one for isolated QDs. Yet, there is a remarkable anomaly.
The emission energy of X$^0_A$ is lower than that of X$^0_B$, indicating that QD A is bigger than QD B. However, the emission energy of
X$^-_A$ is higher than that of X$^-_B$. This result is difficult to explain in terms of isolated QDs. If we assume that the different
structural conditions of the QDs lead to lower energy for X$^0_A$, they should also lead to lower energy for X$^-_A$.\cite{Climente2008}
The anomalous behavior can be explained if the electrons in the QD pair are tunnel coupled. To illustrate this point, in Fig.~\ref{Fig3}(b)
we compare the experimental energies of neutral and negatively charged excitons (left column) with those of a coupled QD pair (right
column). The QDs have the same parameters as in the single QD case of Fig.~\ref{Fig2}, but QD A is now slightly bigger ($\hbar \omega_e=34$
meV). To enable tunnel coupling, we consider the distance between the centers of the QDs is $d=30$ nm, which is slightly below the average
value found by AFM. Under these conditions, not only the calculated energy ordering is the same as in the experiment, but also we obtain
remarkable agreement in the energy splittings. In the experiment, the X$^0$ peaks are split by $1.06$ meV and the X$^-$ peaks by $-1.13$
meV. In the calculations, the corresponding values are $0.96$ meV and $-1.17$ meV, respectively.

\begin{figure}[htb]
\begin{center}
\includegraphics[width=70 mm]{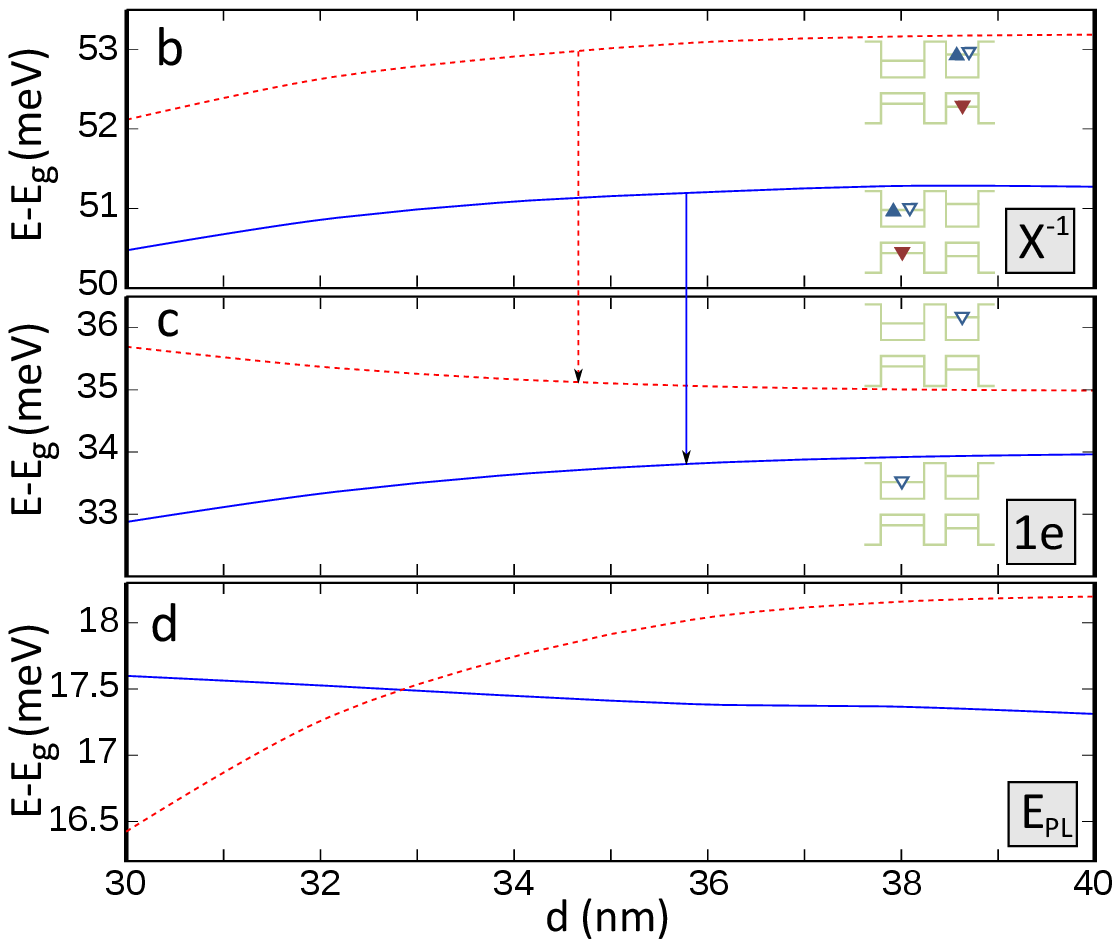}
\caption{B. Al\'en et al.} \label{Fig4}
\end{center}
\end{figure}

To understand the reversed ordering of the trion emission lines as compared to that of excitons, one must notice the different response of
the two species when the QD pair is approached. In asymmetric QD pairs, the energy splitting between the two direct X$^0$ states (X$^0_A$
and X$^0_B$) is barely sensitive to the interdot distance, up to very small separations.\cite{Baira2008} The situation is similar for
X$^-$, as shown in Fig.~\ref{Fig4}(a). However, the final electron states display a pronounced tunnel coupling,\cite{Peng2010a} which
translates into a sizable energy splitting with decreasing interdot distance, see Fig.~\ref{Fig4}(b). In particular, the electron in QD B
evolves towards an antibonding molecular state. As a consequence, the emission energy ($E_{PL}(X^-)=E(X^-)-E(e^-)$) of the trion in QD B
decreases and eventually crosses that of QD A at $d\sim 33$ nm, as shown in Fig.~\ref{Fig4}(c). A further reduction in the interdot
distance results in the reversed ordering of our experimental spectra.

To conclude, we have shown how the number of electrons can be electrically controlled in lateral QD pairs embedded in a lateral Schottky
diode. As the number of confined electrons is lowered, carrier-carrier interactions give rise to well defined energy shifts which we
compare with a theoretical model including Coulomb interactions and electron tunnel coupling. We have found that the negative trion
emission energy is sensitive to single electron tunneling for interdot distances $d<=40$ nm. These results are relevant in the field of
scalable quantum information processing using laterally coupled QDs.

The authors wish to acknowledge to Spanish MICINN through projects Consolider-Ingenio 2010 QOIT (CSD2006-0019), TEC2008-06756-C03-01/03 and
CTQ2008-03344, and to Comunidad Aut\'{o}noma de Madrid through project Q\&CLight (S2009ESP-1503).


\newpage
\Large{\textbf{Figure Captions}}\\

\vspace{6 mm} \small

\textbf{Figure 1.-} (Color online) a), b) and c) show two AFM images and an optical image of a single QD pair, a 1.5-$\mu$m-wide GaAs
channel, and a full MSM device, respectively. d) to f) show contour maps of the bias dependent $\mu$-PL measured in four different
nanostructures. The orange (spotted) lines stand for the $\mu$-PL intensity integrated in the corresponding emission ranges.

\textbf{Figure 2.-} (Color online) a) Electric field \textit{vs.} bias dependence calculated using a semi-analytical transport model [black
line]. Spectrally integrated $\mu$-PL intensity obtained in the same bias range [orange (spotted) line]. b) $\mu$-PL contour plot map
\textit{vs.} lateral bias measured for QN1. The inset compares the experimental [left] and theoretical [right] emission energies for lines
P1-P4. c) Two spectra extracted from the contour plot at the indicated voltages. d) Excitation power evolution of P1-P4 and corresponding
lorentzian fits (solid lines].

\textbf{Figure 3.-} (Color online) a) $\mu$-PL contour plot map \textit{vs.} lateral bias measured for QN2. b) Comparison between
experimental [left] and theoretical [right] energies for X$^0$ and X$^-$ emission from QD A and B ($d=30$ nm).

\textbf{Figure 4.-} (Color online) Calculated energy of the negative trions (a), electrons (b) and corresponding trion emission line (c) as
a function of the interdot distance, $d$. The insets indicate the location of electrons and holes in the long $d$ limit. $E_g$ is the gap
energy.

\newpage

\newpage

\newpage

\newpage

\end{document}